\documentclass[11pt,a4paper,numbers,sort&compress]{article}
\usepackage[utf8]{inputenc}
\pdfoutput=1
\usepackage{jheppub}

\usepackage{graphicx} 
\usepackage{amsmath,amsfonts,amssymb}
\usepackage{physics}
\usepackage{bbm}

\DeclareFontEncoding{LS1}{}{}
\DeclareFontSubstitution{LS1}{stix}{m}{n}

\makeatletter
  \newcommand*\textmathversion{\csname textmv@\math@version\endcsname}
  \newcommand*\textmv@normal{m}
  \newcommand*\textmv@bold{b}
\makeatother

\newcommand\fft[2]{\frac{#1}{#2}}

\newcommand\nn{\nonumber}

\preprint{LCTP-23-14}

\title{Rounding out the story of higher derivative consistent truncations}

\author{James T. Liu and Robert J. Saskowski}
\emailAdd{jimliu@umich.edu, rsaskows@umich.edu}

\affiliation{Leinweber Center for Theoretical Physics, Randall Laboratory of Physics\\The University of Michigan, Ann Arbor, MI 48109-1040, USA }

\abstract{At the two-derivative order, the group manifold reduction of heterotic supergravity on $S^3$ results in a half-maximal 7D gauged supergravity coupled to three vector multiplets, and a further truncation can be taken to remove the vector multiplets.  We demonstrate that this truncation remains consistent at the four-derivative level; we do so both by analysis of the equations of motion and the supersymmetry variations.}
\keywords{}

\date{\today}

\begin{document}
\maketitle

\section{Introduction and summary}
Consistent truncations offer a systematic approach to simplifying various aspects of supergravity theories by selectively eliminating certain degrees of freedom while preserving the underlying symmetries and consistency of the theory.  However, they are also rare and challenging to construct \cite{Freedman:1983zt,Duff:1984hn,Cvetic:2000dm}. Life is simple on a torus $T^n$ since the truncation to the massless sector is a restriction to the $U(1)^n$ singlets. Products of singlets can never yield non-singlets, so symmetry protects the truncation and there is little that can go wrong. However, there are many subtleties involved with sphere truncations (see \textit{e.g.} \cite{deWit:1986oxb,Nastase:1999kf,Lu:1999bc,Cvetic:1999un,Lu:1999bw,Cvetic:1999au,Cvetic:2000dm,Cvetic:2000nc,Lee:2014mla,Nicolai:2011cy,Samtleben:2019zrh,Bonetti:2022gsl,Ciceri:2023bul}) and more general coset reductions (see \textit{e.g.} \cite{Cvetic:2003jy,House:2005yc,Cassani:2009ck,Cassani:2010na,Bena:2010pr}). The issue is that there are fields in the massless sector that transform non-trivially under the isometry group; amongst these is the gauge field that arises from the metric. In such cases, products of the retained fields may act as sources for the very fields we wish to truncate, which spells disaster for consistency. Hence, such truncations are quite delicate.

On the other hand, we are now entering an age of precision holography, where higher-derivative corrections are becoming increasingly important. This is motivated partially by advancements in the construction of higher-derivative supergravities but also by advancements on the field theory side that allow for precise comparison. However, such higher-derivative corrections are subtle. We recently showed that not every two-derivative truncation necessarily leads to a consistent four-derivative one \cite{Liu:2023fqq}. Some higher-derivative gauged supergravities could not be reached by a consistent truncation, in particular, because of non-trivial couplings between the graviton and dilaton-containing multiplets. Hence, it is doubly subtle to check that a four-derivative sphere truncation is indeed consistent.

As in \cite{Liu:2023fqq}, we work in the context of heterotic supergravity with four-derivative corrections. Here, we will be interested in the Scherk-Schwarz reduction on $S^3$, which yields half-maximal 7D gauged supergravity coupled to three vector multiplets. The truncation to pure 7D supergravity is known to be consistent at the two-derivative level \cite{Chamseddine:1999uy}, and, as we will show, it is indeed consistent at the four-derivative level as well.

Our starting point is 10D heterotic supergravity with field content $(g_{MN},\psi_M,B_{MN},\lambda,\phi)$, and leading order bosonic Lagrangian
\begin{equation}
    e^{-1}\mathcal L=e^{-2\phi}\left[R+4(\partial_M\phi)^2-\fft1{12}H_{MNP}^2\right].
\end{equation}
Note that we do not include the heterotic gauge fields, as our aim is to truncate to pure 7D supergravity. For the $S^3$ reduction, we take a metric ansatz
\begin{equation}
        \dd s^2_{10}=g_{\mu\nu}\dd x^\mu\dd x^\nu+g_{ij}\eta^i\eta^j,\qquad \eta^i=\sigma^i+A^i,
\end{equation}
where $x^\mu$ are coordinates on the 7D base space, $A^i$ is a principal $SU(2)$ connection, $\sigma^i$ is a set of left-invariant one-forms on $S^3$, and $g_{ij}$ is a symmetric matrix of scalars. We reduce the two-form as
\begin{equation}
        B=b+B_i\land\eta^i+\frac{1}{2}b_{ij}\,\eta^i\land\eta^j+m\omega_{(2)},
\end{equation}
such that
\begin{equation}
        \dd\omega_{(2)}=\frac{1}{3!}\epsilon_{ijk}\,\sigma^i\land\sigma^j\land\sigma^k,
\end{equation}
is the volume form on $S^3$.  Here, $m$ is the $H_3$ flux supporting the sphere reduction.

At the two-derivative level, this reduction was worked out in \cite{Chamseddine:1999uy}; the resulting theory is a seven-dimensional gauged supergravity with Lagrangian
\begin{equation}
    e^{-1}\mathcal L_7=e^{-2\varphi}\left[R+4(\partial_\mu\varphi)^2-\fft1{12}h_{\mu\nu\lambda}^2-\fft1{2g^2}(F_{\mu\nu}^i)^2+g^2\right],
\end{equation}
where the gauge coupling constant $g$ is related to the flux by $g^2=-1/m$.  We extend this reduction to the four-derivative level.  In particular, after analyzing the bosonic equations of motion, we find a consistent $\mathcal O(\alpha')$ truncation
\begin{align}
    g_{ij}=\fft1{\tilde g^2}\delta_{ij}+\frac{\alpha'}{4\tilde g^4}F^i_{\alpha\beta}F^j_{\alpha\beta},\qquad b_{ij}=0,\qquad B_i=-\fft1{\tilde g^2}A^i.
\end{align}
where the gauge coupling constant, $\tilde g$, receives an $\mathcal O(\alpha')$ shift
\begin{equation}
    \fft1{\tilde g^2}=\fft1{g^2}-\fft{\alpha'}2=-m-\fft{\alpha'}2,
\end{equation}
Note that this truncation reduces to the torus case of \cite{Liu:2023fqq} in the $g\to0$ limit, assuming that the fields are appropriately rescaled.

We also consider the fermionic sector. Here, the 10D gravitino $\psi_M$ splits into a 7D gravitino $\psi_\mu$ and three gaugini $\psi_i$. Note that the left-invariant one-form basis trivializes the spin bundle and hence the reduction preserves the full supersymmetry of our original theory. In order to truncate the gaugini, we require a field redefinition
\begin{equation}
    \tilde\psi_i=\psi_i-\frac{\alpha'}{2\tilde g}F^i_{\mu\nu}\mathcal D_\mu\psi_\nu,
\end{equation}
where we have denoted $\delta_\epsilon\psi_\mu=\mathcal D_\mu\epsilon$. This then sets $\delta_\epsilon\tilde\psi_i=0$ and leads to consistency with supersymmetry.

The rest of this note is organized as follows. In Section \ref{sec:reduction}, we review the two-derivative group manifold reduction of heterotic supergravity on $S^3$. We then show, in Section \ref{sec:bosonic}, that the truncation is consistent at the four-derivative level by analysis of the equations of motion, as well as summarize the remaining equations of motion and effective Lagrangian. In Section \ref{sec:fermionic}, we check that this truncation is consistent with supersymmetry, and, finally, we conclude in Section \ref{sec:discussion}.

\section{Group manifold reduction on $S^3$}\label{sec:reduction}

The reduction of heterotic supergravity on $S^3$ can be viewed as a reduction on the group manifold of $SU(2)$.  To introduce our notation and conventions, we first summarize the standard results of the group manifold reduction \cite{Chamseddine:1999uy,Cvetic:2003jy,Lu:2006ah}. As a matter of setting notation, we will use early capital Latin letters for 10D rigid indices and late capital Latin letters for 10D curved indices. Our index splitting convention is summarized as
\begin{equation*}
    M\to \{\mu,i\},\qquad A\to \{\alpha, a\}.
\end{equation*}
We use $\hat\nabla$ to mean the Levi-Civita connection in 10D, while we use $\nabla$ for the Levi-Civita connection on the base space.

\subsection{Four-derivative heterotic supergravity}

In the string frame, the ten-dimensional bosonic Lagrangian up to four-derivative corrections is given by \cite{Bergshoeff:1988nn,Bergshoeff:1989de,Metsaev:1987zx,Chemissany:2007he}
\begin{equation}
    e^{-1}\mathcal L=e^{-2\phi}\left[R+4(\partial_M\phi)^2-\fft1{12}\tilde H_{MNP}^2+\frac{\alpha'}{8}\big(R_{MNAB}(\Omega_+)\big)^2\right]+\mathcal{O}(\alpha'^3),
\label{eq:Lhet}
\end{equation}
where $R$ is the Ricci scalar and we have defined
\begin{equation}
    \tilde H= H-\frac{\alpha'}{4}\omega_{3L}(\Omega_+),
\label{eq:Htilde}
\end{equation}
where $H=\dd B$ is the three-form flux. Note that we have truncated out the heterotic gauge fields, as they will not play an important role in our discussion.  Here we have introduced the torsionful connection
\begin{equation}
    \Omega_\pm=\Omega\pm\fft12\mathcal H,\qquad\mathcal H^{AB}\equiv \tilde H_M{}^{AB}\dd x^M,
\end{equation}
where $\Omega$ is the ordinary ten-dimensional spin connection.  The torsionful Riemann tensor and Lorentz Chern-Simons form are defined as usual
\begin{align}
    R(\Omega_\pm)&=\dd\Omega_\pm+\Omega_\pm\wedge\Omega_\pm,\nn\\
    \omega_{3L}(\Omega_\pm)&=\Tr\left(\Omega_\pm\wedge \dd\Omega_\pm+\fft23\Omega_\pm\wedge\Omega_\pm\wedge\Omega_\pm\right).
\end{align}
The interplay between $\Omega_+$ and $\Omega_-$ plays an important role in supersymmetry \cite{Bergshoeff:1988nn,Bergshoeff:1989de}.

As in \cite{Liu:2023fqq}, we break up the bosonic equations of motion into two- and four-derivative parts
\begin{align}
    \mathcal E_\phi^{(0)}&=R-4(\partial_M\phi)^2+4\hat\Box\phi-\fft1{12} H_{MNP}^2,\nn\\
    \mathcal E_{g,MN}^{(0)}&=R_{MN}+2\hat\nabla_M\hat\nabla_N\phi-\fft14 H_{MAB} H_N{}^{AB},\nn\\
    \mathcal E_{H,NP}^{(0)}&=e^{2\phi}\hat\nabla^M\qty(e^{-2\phi} H_{MNP}),
\label{eq:eom2d}
\end{align}
and
\begin{align}
    \mathcal E_\phi^{(1)}&=\fft1{24}H_{MNP}\omega_{3L}{}^{MNP}(\Omega_+)+\fft18\qty(R_{MNAB}(\Omega_+))^2,\nn\\
    \mathcal E_{g,MN}^{(1)}&=\fft18 H_{MAB} \omega_{3L\,N}{}^{AB}(\Omega_+)+\fft14R_{MPAB}(\Omega_+)R_N{}^{PAB}(\Omega_+),\nn\\
    \mathcal E_{H,NP}^{(1)}&=-\fft14e^{2\phi}\hat\nabla^M\qty(e^{-2\phi} \omega_{3L,MNP}(\Omega_+)).
\label{eq:eom4d}
\end{align}
Likewise, the supersymmetry variations have two- and four-derivative pieces
\begin{align}
    \delta_\epsilon\psi_M^{(0)}&=\left(\nabla_M-\fft18H_{MNP}\Gamma^{NP}\right)\epsilon,&\delta_\epsilon\psi_M^{(1)}&=\fft1{32}\omega_{3L,MNP}(\Omega_+)\Gamma^{NP}\epsilon,\nn\\
    \delta_\epsilon\lambda^{(0)}&=\left(\Gamma^M\partial_M\phi-\fft1{12} H_{MNP}\Gamma^{MNP}\right)\epsilon,&\delta_\epsilon\lambda^{(1)}&=\fft1{48}\omega_{3L,MNP}(\Omega_+)\Gamma^{MNP}\epsilon.
\label{eq:deltas}
\end{align}
We note that our conventions for 10D heterotic supergravity are identical to those of \cite{Liu:2023fqq}.

\subsection{$S^3$ reduction}
We now proceed to reduce the heterotic theory on $S^3$ to arrive at half-maximal seven-dimensional gauged supergravity coupled to three vector multiplets \cite{Chamseddine:1999uy,Lu:2006ah}.  The vector multiplets will be truncated out in the next section, but we retain them here since we will need to make use of the full non-truncated lowest-order equations of motion in order to compute their four-derivative shifts.

The sphere $S^3$ is, as a manifold, isomorphic to $SU(2)$, which admits a basis of globally defined left-invariant one-forms $\sigma^i$ satisfying the Maurer-Cartan equation
\begin{equation}
        \dd \sigma^i=-\frac{1}{2}\epsilon^{ijk}\sigma^j\land\sigma^k.
\end{equation}
Such a global frame reduces the structure group to the identity, which is just the statement that $S^3$ is parallelizable. Moreover, $\sigma^i$ generates the right isometries of the metric. The $\epsilon^{ijk}$ are the structure constants of $\mathfrak{su}(2)$, and formally the indices should be raised and lowered via the Cartan-Killing metric $\kappa$. However, in this case, the Cartan-Killing metric is trivial
\begin{equation}
        \kappa_{ij}\equiv -\frac{1}{2}\epsilon^{k}{}_{\ell i}\epsilon^{\ell}{}_{kj}=\delta_{ij},
\end{equation}
and so we will not concern ourselves with the positions of the indices of $\epsilon$. 
    
We then have a metric ansatz in the form of a standard Scherk-Schwarz reduction \cite{Scherk:1979zr}
\begin{equation}
        \dd s^2_{10}=g_{\mu\nu}\dd x^\mu\dd x^\nu+g_{ij}\eta^i\eta^j,\qquad \eta^i=\sigma^i+A^i,\label{eq:metric}
\end{equation}
where $x^\mu$ are coordinates on the 7D base space, $g_{ij}$ is a symmetric matrix of scalars and $A^i$ forms a principal $SU(2)$ connection with curvature locally given by
\begin{equation}
        F^i=\dd A^i-\frac{1}{2}\epsilon^{ijk}A^j\land A^k.
\end{equation}
Being non-abelian, our gauge field naturally has an associated gauge-covariant derivative, which we will denote as $D$. Given an $\mathfrak{su}(2)$-valued form $t^i$, $D$ acts as
\begin{equation}
        Dt^i=\dd t^i-\epsilon^{ijk}A^j\land t^k.
\end{equation}
Considering the metric \eqref{eq:metric}, we choose a natural zehnbein
\begin{equation}
        E^\alpha=e^\alpha,\qquad E^a=e^a_i\eta^i.
\end{equation}
where $e^\alpha$ is a siebenbein for $g_{\mu\nu}$ and $e^a$ is a dreibein for $g_{ij}$, so that $\dd s_{10}^2=\eta_{\alpha\beta}E^\alpha E^\beta+\delta_{ab}E^aE^b$. We then compute
\begin{align}
        \dd E^\alpha&=-\omega^{\alpha\beta}E^\beta,\nn\\
        \dd E^a&=\frac{1}{2}e_i^aF^i_{\alpha\beta}E^\alpha\land E^\beta+e^i_b D_\alpha e_i^a E^\alpha\land E^b-\frac{1}{2}\epsilon^{ijk}e_i^ae_b^je_c^k E^b\land E^c.
\end{align}
from which one may extract the components of the spin connection
\begin{align}
        \Omega^{\alpha\beta}&=\omega^{\alpha\beta}-\frac{1}{2}e_i^a F_{\alpha\beta}^i E^a,\nn\\
        \Omega^{\alpha b}&=-P_\alpha^{bc}E^c-\frac{1}{2}e_i^b F^i_{\alpha\beta}E^\beta,\nn\\
        \Omega^{ab}&=Q_\alpha^{ab} E^\alpha+\frac{1}{2}\mathcal C_{c,ab}E^c.
\end{align}
Here, we have made the conventional definitions
\begin{align}
        P_\alpha^{ab}&=e^{i(a} D_\alpha e^{b)}_i,\nn\\
        Q_\alpha^{ab}&=e^{i[a} D_\alpha e^{b]}_i,\nn\\
        \mathcal C_{c,ab}&=\epsilon^{ijk}\qty[e^c_i e^j_a e^k_b+e^b_i e^j_a e^k_c-e_i^a e^j_b e^k_c],
\end{align}
where $P$ and $Q$ are the scalar kinetic term and composite $SU(2)$ connection, respectively.

We may then compute the relevant Riemann tensor components to be
\begin{align}
        R_{\gamma\delta}{}^{\alpha\beta}(\Omega)=&R^{\alpha\beta}{}_{\gamma\delta}(\omega)-\frac{1}{2}F^i_{\alpha\beta}F^j_{\gamma\delta}g_{ij}-\frac{1}{2}F^i_{\alpha\gamma}F^j_{\beta\delta}g_{ij},\nn\\
        R_{\gamma d}{}^{\alpha\beta}(\Omega)=&-\frac{1}{2}D_\gamma\qty(e^d_i F_{\alpha\beta}^i)-\frac{1}{2}e_i^a F_{\alpha\beta}^i\qty(e^j_d D_\gamma e^a_j)-e^c_i F_{\alpha\gamma}^iP_{\beta cd},\nn\\
        R_{c\delta}{}^{\alpha b}(\Omega)=&D_\delta P_{\alpha bc}+P_{\alpha bd}e^i_c D_\delta e_i^d+\frac{1}{4}e_i^c e_j^b F_{\alpha\gamma}^iF_{\gamma\delta}^j-P_{\alpha dc}Q_{\delta db}+\frac{1}{4}e_i^d F^i_{\alpha\delta}\mathcal C_{c,db},\nn\\
        R_{cd}{}^{\alpha b}(\Omega)=&P_{\alpha ba}\epsilon^{ijk}e_i^ae^j_ce^k_d+e_i^cF_{\alpha\gamma}^iP_{\gamma bd}-P_{\alpha ac}\mathcal C_{d,ab},\nn\\
        R_{cd}{}^{ab}(\Omega)=&-2P_{\gamma ac}P_{\gamma bd}-\frac{1}{4}\mathcal C_{f,ab}\epsilon^{ijk}e_i^fe_j^ce_k^d+\frac{1}{2}\mathcal C_{c,af}\mathcal C_{d,fb},
\end{align}
where there is implicit antisymmetrization as relevant. Contracting Riemann then gives the Ricci tensor components
\begin{align}
        R_{\alpha\beta}(\Omega)=&R_{\alpha\beta}(\omega)-\frac{1}{2}F_{\alpha\gamma}^iF_{\beta\gamma}^j g_{ij}-D_\beta P_{\alpha cc}-P_{\alpha cd}P_{\beta}{}^{cd},\nn\\
        R_{\alpha b}(\Omega)=&\frac{1}{2}D_{\gamma}\qty(e_i^b F_{\alpha\gamma}^i)+\frac{1}{2}e_i^aF_{\alpha\gamma}^ie^j_bD_\gamma e_j^a+\frac{1}{2}e_i^bF_{\alpha\gamma}^iP_{\gamma cc},\nn\\
        R_{a b}(\Omega)=&-D^\gamma P_{\gamma ab}-P_{\gamma ad}\qty(e^i_bD_\gamma e^d_i)+\frac{1}{4}e_i^ae_j^bF_{\gamma\delta}^iF_{\gamma\delta}^j+P_{\gamma db}Q_{\gamma da}-2P_{\gamma a[b|}P_{\gamma c|c]}\nonumber\\
        &-\frac{1}{4}\mathcal C_{f,ac}\epsilon^{ijk}e_i^fe_j^be_k^c,
\end{align}
and the Ricci scalar
\begin{equation}
        R(\Omega)=R(\omega)-\frac{1}{4}F^i_{\alpha\beta}F^j_{\alpha\beta}g_{ij}-2\nabla^\gamma P_{\gamma cc}-P_{\gamma cc}P^{\gamma dd}-P^2-\frac{1}{4}\qty(\epsilon^{ijk}\epsilon^{\ell mn}h_{i\ell}h^{jm}h^{kn}+2\epsilon^{ijk}\epsilon^{ji\ell}h^{k\ell}).
\end{equation}

For the reduction, we also need to consider the dilaton and $B$-field.  For derivatives of the dilaton, we find
\begin{equation}
    \hat\nabla_\alpha\hat\nabla_\beta\phi=\nabla_\alpha\nabla_\beta\phi,\qquad\hat\nabla_\alpha\hat\nabla_b\phi=-\frac{1}{2}e_{ib}F^{i}_{\alpha\gamma}\partial^\gamma\phi,\qquad \hat\nabla_a\hat\nabla_b\phi=P_{\gamma ab}\partial^\gamma\phi.
\end{equation}
For the $B$-field, we make the sphere reduction ansatz
\begin{equation}
        B=b+B_i\land\eta^i+\frac{1}{2}b_{ij}\,\eta^i\land\eta^j+m\omega_{(2)},
\end{equation}
such that
\begin{equation}
        \dd\omega_{(2)}=\frac{1}{3!}\epsilon_{ijk}\,\sigma^i\land\sigma^j\land\sigma^k,
\end{equation}
is the volume form on $S^3$ and $m$ is the three-form flux supporting the reduction.  This then leads to an expression for the three-form field strength in terms of the seven-dimensional field strengths (three-form $h$, two-form $\tilde G_i$ and one-form $G_{ij}$) according to
\begin{equation}
        H=h+\tilde G_i\land\eta^i+\frac{1}{2}G_{ij}\land\eta^i\land\eta^j+\frac{m}{6}\epsilon_{ijk}\,\eta^i\land\eta^j\land\eta^k,
\end{equation}
where
\begin{align}
        h&=\dd b-B_i\land F^i-\frac{m}{6}\epsilon_{ijk}A^i\land A^j\land A^k,\nn\\
        \tilde G_i&=G_i-b_{ij}F^j+\frac{m}{2}\epsilon_{ijk}A^j\land A^k,\qquad G_i=D B_i,\nn\\
        G_{ij}&=Db_{ij}+\epsilon_{kij}B_k-m\epsilon_{kij}A^k.
\label{eq:hGiGij}
\end{align}
Note that these field strengths satisfy the Bianchi identities
\begin{align}
        \dd h&=-\tilde G_i\land F^i,\nn\\
        D\tilde G_i&=-G_{ij}\land F^j,\nn\\
        DG_{ij}&=-m\epsilon_{ijk}F^k+\epsilon_{kij}\tilde G_k.\label{eq:ogBianchi}
\end{align}

\subsection{The bosonic reduction at leading order}
At the leading two-derivative level, the above reduction ansatz leads to the reduced bosonic Lagrangian
\begin{align}
        e^{-1}\mathcal L^{(0)}&=e^{-2\varphi}\Bigl[R+4(\partial\varphi)^2-\frac{1}{12}h^2-\frac{1}{4}\qty(g_{ij}F^i_{\alpha\beta}F^j_{\alpha\beta}+g^{ij}\tilde G_{\mu\nu\,i}\tilde G_{\mu\nu\,j})\nn\\
        &\kern4em-(P_\mu{}^{ab})^2-\frac{1}{4}g^{ij}g^{k\ell}G_{\mu ik}G_{\mu j\ell}\nn\\
        &\kern4em-\frac{1}{4}\qty(\epsilon^{ijk}\epsilon^{\ell mn}g_{i\ell}g^{jm}g^{kn}+2\epsilon^{ijk}\epsilon^{ji\ell}g^{k\ell})-\frac{1}{2}m^2\Bigr].\label{eq:redL}
\end{align}
For comparison, note that the first two lines of \eqref{eq:redL} is a gauge covariantized version of a standard torus reduction \cite{Maharana:1992my,Liu:2023fqq}, while the last line is a scalar potential generated by the gauged geometry. Note also that there is a natural identification of the 7D dilaton as
\begin{equation}
        \varphi=\phi-\frac{1}{4}\log\det g_{ij}.\label{eq:dilatonshift}
\end{equation}
Along with the reduced Lagrangian, we have a set of reduced equations of motion at leading order.  The reduced Einstein equation becomes
\begin{align}\label{eq:redEinstein}
        \mathcal E^{(0)}_{g,\alpha\beta}=\,&R_{\alpha\beta}(\omega)-\frac{1}{2}\qty(F_{\alpha\gamma}^iF_{\beta\gamma}^jg_{ij}+\tilde G_{\alpha\gamma i}\tilde G_{\beta\gamma j}g^{ij})+2\nabla_\alpha\nabla_\beta\varphi-P_{\alpha cd}P_\beta{}^{cd}-\frac{1}{4}h_{\alpha\gamma\delta}h_\beta{}^{\gamma\delta}\nonumber\\
        &-\frac{1}{4}G_{\alpha ij}G_{\beta k\ell}g^{ik}g^{j\ell},\nn\\
        \mathcal E^{(0)}_{g,\alpha b}=\,&\frac{1}{2}e^{2\varphi}D^{\gamma}\qty(e^{-2\varphi}F_{\alpha\gamma}^i)e_i^b+F_{\alpha\gamma}^ie_i^a P_{\gamma ba}-\frac{1}{4}h_{\alpha\gamma\delta}\tilde G_i^{\gamma\delta}e^i_b+\frac{1}{2}\tilde G_{\alpha\gamma i}G^{\gamma}_{jk}g^{ik}e^j_b\nonumber\\
        &-\frac{m}{4}G_{\alpha ij}g^{i\ell}g^{jm}\epsilon_{k\ell m}e^k_b,\nn\\
        \mathcal E^{(0)}_{g,ab}=\,&-e^{2\varphi}\nabla^\gamma (e^{-2\varphi} P_{\gamma ab})+2P_{\gamma (a|d}Q_{\gamma d|b)}-\frac{1}{12}\mathcal C_{f,ac}\mathcal C_{f,bc}\nonumber\\
        &+\frac{1}{4}\qty(F_{\gamma\delta}^iF_{\gamma\delta}^je_i^ae_j^b-\tilde G_{\gamma\delta i}\tilde G_{\gamma\delta j}e^i_a e^j_b)-\frac{1}{2}G_{\delta ik}G_{\delta j\ell}h^{k\ell}e^i_a e^j_b\nonumber\\
        &-\frac{m^2}{4}\epsilon_{ik\ell}\epsilon_{jmn}g^{km}g^{\ell n}e^i_a e^j_b,
\end{align}
while the $H$ equation becomes
\begin{align}\label{eq:redH}
        \mathcal E^{(0)}_{H,\alpha\beta}=\,&e^{2\varphi}\nabla^\gamma\qty(e^{-2\varphi}h_{\gamma\alpha\beta}),\nn\\
        \mathcal E^{(0)}_{H,\alpha b}=\,&e^{2\varphi}D^\gamma\qty(e^{-2\varphi}\tilde G_{\gamma\alpha i})e^i_b-2P_{\gamma bd}\tilde G_{\gamma\alpha i}e^i_d+\frac{1}{2}h_{\alpha\gamma\delta}F^i_{\gamma\delta}e_i^b+\frac{1}{2}\mathcal C_{c,bd}G_{\alpha ij}e^i_d e^j_c,\nn\\
        \mathcal E^{(0)}_{H,ab}=\,&e^{2\varphi}\nabla^\mu\qty(e^{-2\varphi}G_{\mu ij}e^i_a e^j_b)-F_{\alpha\gamma}^i\tilde G_{\alpha\gamma j}e_i^{[a}e^{b]j}+2Q_{\alpha[a|c} G_{\alpha c|b]}\nonumber\\
        &-2P_{\gamma c[a|}G_{\gamma ij}e^i_ce^j_{|b]}-\mathcal C_{c,[a|d}\epsilon^{ijk}e^i_d e^j_c e^k_{|b]}.
\end{align}
Finally, the dilaton equation becomes
\begin{align}\label{eq:redPhi}
        \mathcal E^{(0)}_{\phi}&=R(\omega)-\frac{1}{4}\qty(F^i_{\alpha\beta}F^j_{\alpha\beta}g_{ij}+\tilde G_{\alpha\beta i}\tilde G_{\alpha\beta j}g^{ij})-\frac{1}{12}h^2-4(\partial\varphi)^2+4\Box\varphi-\qty(P_{\mu ab})^2\nonumber\\
        &-\frac{1}{4}G_{\alpha ij}G_{\alpha k\ell}g^{ik}g^{j\ell}-\frac{1}{12}\mathcal C^2-\frac{m^2}{12}\epsilon^{ijk}\epsilon^{\ell mn}g^{i\ell}g^{jm}g^{kn}.
\end{align}
It is straightforward to check that the above equations of motion follow from the Lagrangian \eqref{eq:redL}, which confirms that the reduction is indeed consistent.
    
\subsection{The fermionic reduction at leading order}
We may also reduce the variations of the fermion fields. Since the dilaton is shifted \eqref{eq:dilatonshift}, we must likewise shift the dilatino as
\begin{equation}
        \tilde\lambda=\lambda-\Gamma^i\psi_i.
\end{equation}
The leading order fermion transformations then reduce to
\begin{align}
        \delta_\epsilon\psi_\mu^{(0)}&=\qty[\nabla_\mu(\omega_-)+\frac{1}{4}Q_{\mu ab}\Gamma^{ab}+\frac{1}{4}\qty(g_{ij}F^j_{\mu\nu}-\tilde G_{\mu\nu i})\gamma^\nu\Gamma^i-\frac{1}{8}G_{\mu ij}\Gamma^{ij}]\epsilon,\nn\\
        \delta_\epsilon\psi_i^{(0)}&=\Bigg[-\frac{1}{8}\qty(g_{ij}F^j_{\mu\nu}+\tilde G_{\mu\nu i})\gamma^{\mu\nu}-\frac{1}{2}e_i^b\qty(P_{\mu ab}+\frac{1}{2}G_{\mu jk}e_a^je^k_b)\gamma^\mu \Gamma^a\nonumber\\
        &\qquad+\frac{1}{8}\qty(\mathcal C_{c,ab}e^c_i-m\epsilon_{ijk}e^j_a e^k_b)\Gamma^{ab}\Bigg]\epsilon,\nn\\
        \delta_\epsilon\tilde\lambda^{(0)}&=\qty[\gamma^\mu\partial_\mu\varphi-\frac{1}{12}h_{\mu\nu\rho}\gamma^{\mu\nu\rho}+\frac{1}{8}\qty(g_{ij}F^j_{\mu\nu}-\tilde G_{\mu\nu i})\gamma^{\mu\nu}\Gamma^i-\frac{1}{8}\qty(\mathcal C_{c,ab}-\frac{m}{3}\epsilon_{ijk}e^i_ce^j_a e^k_b)\Gamma^{abc}]\epsilon.\label{eq:redVar}
    \end{align}
Notice that the composite connection $Q$ appears in the gravitino variation to make the derivative covariant with respect to this connection.  Since the reduced gravitino, $\psi_\mu$, and dilatino, $\tilde\lambda$, are in the gravity multiplet, while the internal gravitino components, $\psi_i$, are in the vector multiplets, we can identify the graviphoton and vector multiplet gauge field combinations as
\begin{align}
        F_{\mu\nu}^{a\,(-)}&=e^a_iF_{\mu\nu}^i-e_a^i\tilde G_{\mu\nu\,i},&&(\hbox{graviphoton})\nn\\
        F_{\mu\nu}^{a\,(+)}&=e^a_iF_{\mu\nu}^i+e_a^i\tilde G_{\mu\nu\,i}.&&(\hbox{vector})
    \label{eq:gpv}
\end{align}
This matches the torus case of \cite{Liu:2023fqq}, and will be used as a guide to truncating out the three vector multiplets below.

\section{The bosonic truncation}\label{sec:bosonic}

While the SU(2) reduction includes three additional vectors coming from the reduced $B$-field, it is possible to consistently truncate them away at the two-derivative level, leading to pure 7D gauged supergravity \cite{Chamseddine:1999uy,Lu:2006ah}.  In this section, we demonstrate that it remains consistent to truncate out the additional vector multiplets at the four-derivative level by analysis of the bosonic equations of motion.  Before doing so, however, we review how the truncation works at the two-derivative level in order to set the stage for the four-derivative truncation.

\subsection{The leading order truncation}
At the leading order, the natural choice of truncation is \cite{Chamseddine:1999uy,Lu:2006ah}
\begin{equation}
    g_{ij}=g^{-2}\delta_{ij},\qquad b_{ij}=0,\qquad B_{i}=-g_{ij}A^j=-g^{-2}A^i,
\label{eq:truncation}
\end{equation}
where we have introduced the 7D gauge coupling constant $g$.  Here the choice of minus sign in the relation between $B_i$ and $A^i$ is motivated by the desire to truncate away the vector multiplets as identified in \eqref{eq:gpv}.  As expected for a sphere reduction, the gauge coupling $g$ is necessarily related to the three-form flux $m$ on $S^3$.  To fix the relation between $m$ and $g$, we note that, since $B_i\propto A^i$, we expect a similar relation with the field strengths, $\tilde G_i\propto F^i$.  Substituting (\ref{eq:truncation}) into (\ref{eq:hGiGij}), we obtain
\begin{equation}
    \tilde G_i=-g^{-2}\qty[\dd A^i-\qty(1+\frac{m g^2}{2})\epsilon^{ijk}A^j\land A^k],
\end{equation}
and so we must pick $m=-g^{-2}$ in order to get a properly covariant field strength, $\tilde G_i=-g_{ij}F^i$. This can also be seen from the truncated scalar field strength term
\begin{equation}
    G_{ij}=-\epsilon_{ijk}(m+g^{-2})A^k.
\end{equation}
We expect this expression to vanish since we are truncating away the scalars with $b_{ij}=0$.

Alternatively, we could have started by freezing out the scalars with
\begin{equation}
    g_{ij}=g^{-2}\delta_{ij},\qquad b_{ij}=0,
\end{equation}
in which case the scalar equation arising from the internal Einstein equation becomes
\begin{equation}
        \mathcal E^{(0)}_{g,ab}=\frac{1}{4}\qty(g^{-2}F^i_{\gamma\delta}F^j_{\gamma\delta}-g^2\tilde G_{i\gamma\delta}\tilde G_{j\gamma\delta})\delta^i_a\delta^j_b-\frac{g^6}{2}\qty(m^2-g^{-4})\delta_{ab}.
\end{equation}
This tells us that we must pick the truncation
\begin{equation}
    B_i=\pm g^{-2}A^i,\qquad m=\pm g^{-2},
\end{equation}
to consistently remove this scalar equation as a constraint.  At the bosonic two-derivative level, either sign choice is valid, suggesting that either the graviphotons or the vector multiplet vectors can be removed.  However, based on supersymmetry, we must choose the minus sign to truncate out the vector multiplets, while preserving supersymmetry in the gravity multiplet.

After truncation, the two-derivative equations of motion become
\begin{align}
    \mathcal E^{(0)}_{g,\alpha\beta}&=R_{\alpha\beta}(\omega)-g^{-2}F^i_{\alpha\gamma}F^i_{\beta\gamma}+2\nabla_\alpha\nabla_\beta\varphi-\frac{1}{4}h_{\alpha\gamma\delta}h_\beta{}^{\gamma\delta},\nn\\
    \mathcal E^{(0)}_{g,\alpha b}&=\frac{1}{2g}\qty[e^{2\varphi}D^\gamma\qty(e^{-2\varphi}F^i_{\gamma\alpha})-\frac{1}{2}h_{\alpha\gamma\delta}F^i_{\gamma\delta}]\delta^i_b,\nn\\
    \mathcal E^{(0)}_{g,ab}&=0,\nn\\
    \mathcal E^{(0)}_{H,\alpha\beta}&=e^{2\varphi}\nabla^\gamma\qty(e^{-2\varphi}h_{\gamma\alpha\beta}),\nn\\
    \mathcal E^{(0)}_{H,\alpha b}&=-g^{-1}\qty[e^{2\varphi}D^\gamma\qty(e^{-2\varphi}F^i_{\gamma\alpha})-\frac{1}{2}h_{\alpha\gamma\delta}F^i_{\gamma\delta}]\delta^i_b,\nn\\
    \mathcal E^{(0)}_{H,ab}&=0,\nn\\
    \mathcal E^{(0)}_{\varphi}&=R(\omega)-\frac{1}{2g^2}F^i_{\alpha\beta}F^i_{\alpha\beta}-\frac{1}{12}h^2-4(\partial\varphi)^2+4\Box\varphi+g^2.\label{eq:twoderivEOMs}
\end{align}
In particular, the scalar equations vanish and the gauge field equations are proportional; hence, the truncation is indeed consistent. Note also that, after truncation, the $h$ Bianchi identity becomes
\begin{align}
        \dd h&=g^{-2}F_i\land F^i.\label{eq:hbianchi}
    \end{align}
The above equations of motion, \eqref{eq:twoderivEOMs}, correspond to the reduced Lagrangian \cite{Chamseddine:1999uy,Lu:2006ah}
\begin{align}
        e^{-1}\mathcal L^{(0)}=&e^{-2\varphi}\qty[R+4(\partial\varphi)^2-\frac{1}{12}h^2-\frac{1}{2g^2}\qty(F^i_{\alpha\beta})^2+g^2],
\end{align}
which matches the bosonic sector of gauged half-maximal 7D supergravity \cite{Townsend:1983kk} with gauge coupling constant $g$ related to the flux on $S^3$ according to $g^2=-1/m$.  (In our conventions, this indicates that the flux parameter $m$ is negative.)

\subsection{The truncation at $\mathcal O(\alpha')$}

We now extend the truncation at the four-derivative level.  Here it is important to note that the two-derivative truncation, (\ref{eq:truncation}), may require $\mathcal O(\alpha')$ corrections.  We thus write
\begin{align}
    &B_{\mu i}=-g^{-2}A_\mu^{i}+\alpha'\delta B_{\mu i}\qquad
    g_{ij}=g^{-2}\delta_{ij}+\alpha'\delta g_{ij},\qquad b_{ij}=0+\alpha'\delta b_{ij}.
\label{eq:corr}
\end{align}
We also split the equations of motion as
\begin{equation}
        \mathcal E=\mathcal E^{(0)}+\alpha'\qty(\delta\mathcal E^{(0)}+\mathcal E^{(1)}),
\end{equation}
where $\delta \mathcal E^{(0)}$ is the shift of $\mathcal E^{(0)}$ generated by the field redefinitions \eqref{eq:corr}. At the order we are interested in, the four-derivative piece $\mathcal E^{(1)}$ will only depend on the two-derivative truncation. As in the torus reduction case \cite{Liu:2023fqq}, the necessary condition for consistency of the truncation is that the scalar equations vanish
\begin{equation}
        \delta\mathcal E^{(0)}_{g,ij}+\mathcal E^{(1)}_{g,ij}=0,\qquad 
        \delta\mathcal E^{(0)}_{H,ij}+\mathcal E^{(1)}_{H,ij}=0,
\end{equation}
and the vector equations are compatible
\begin{equation}
        \delta\mathcal E^{(0)}_{g,\alpha i}+\mathcal E^{(1)}_{g,\alpha i}=-\frac{1}{2}\qty(\delta\mathcal E^{(0)}_{H,\alpha i}+\mathcal E^{(1)}_{H,\alpha i}).
\end{equation}

In order to compute the $\mathcal O(\alpha')$ equations of motion, $\mathcal E^{(1)}$, we require the torsionful Riemann tensor $R_{MN}{}^{AB}(\Omega_+)$ and the Lorentz-Chern-Simons form $\omega_{3L}(\Omega_+)$.  These can be obtained using the lowest order torsionful spin connection components, which become, after truncation
\begin{equation}
        \Omega_+=\begin{pmatrix}
        \omega_+^{\alpha\beta}-g^{-2}F_{\alpha\beta}^i\eta^i&0\\
        0&-\epsilon^{ijk}\delta^i_a\delta^j_b\sigma^k
        \end{pmatrix}.
\end{equation}
This then leads to expressions for the torsionful Riemann tensor components
\begin{align}
        R_{\gamma\delta}{}^{\alpha\beta}(\Omega_+)=&R_{\gamma\delta}{}^{\alpha\beta}(\omega_+)-g^{-2} F^i_{\alpha\beta}F^i_{\gamma\delta},\nn\\
        R_{\gamma d}{}^{\alpha\beta}(\Omega_+)=&-g^{-1}D_\gamma^{(+)}F_{\alpha\beta}^i\delta_i^d,\nn\\
        R_{cd}{}^{\alpha\beta}(\Omega_+)=&2g^{-2} F_{\alpha\gamma}^iF_{\gamma\beta}^j\delta_i^{[c}\delta_j^{d]}+\epsilon^{ijk}F^i_{\alpha\beta}\delta^j_c\delta^k_d,&
\end{align}
with all other independent components vanishing. Here $D^{(+)}$ is taken to mean $D(\omega_+)$. The Lorentz-Chern-Simons form is given by
\begin{align}
    \omega_{3L,\alpha\beta\gamma}(\Omega_+)&=\omega_{3L,\alpha\beta\gamma}(\omega_+)+2\epsilon^{ijk}A^i_\alpha A^j_\beta A^k_\gamma,\nonumber\\
    \omega_{3L,\alpha\beta c}(\Omega_+)&=\delta^i_c\qty(2g^{-1} R_{\alpha\beta}{}^{\gamma\delta}(\omega_+)F_{\gamma\delta}^{i}-g^{-3} F_{\alpha\beta}^{j}F_{\gamma\delta}^{j}F_{\gamma\delta}^{i}-2\epsilon^{ijk}A^j_\alpha A^k_\beta),\nonumber\\
    \omega_{3L,\alpha bc}(\Omega_+)&=\delta^{[i}_b\delta^{j]}_c\qty(2g^{-2}F_{\gamma\delta}^{i}D_\alpha^{(+)}F_{\gamma\delta}^{j}+2\epsilon^{kij}A^k_\alpha),\nonumber\\
    \omega_{3L,abc}(\Omega_+)&=\delta_a^{[i}\delta_b^j\delta_c^{k]}\qty(-4g^{-3}F_{\alpha\beta}^{i}F_{\beta\gamma}^{j}F_{\gamma\alpha}^{k}-2\epsilon^{ijk}).
\label{eq:LCSf}
\end{align}
Given the shifted $H$-field, \eqref{eq:Htilde}, the additional terms proportional to $A^i$ in $\omega_{3L}(\Omega_+)$ hint that we should define modified field strengths
\begin{align}
        \bar h&=\dd b-B_i\land F^i+\frac{1}{6}\qty(g^{-2}-\frac{\alpha'}{2})\epsilon_{ijk}A^i\land A^j\land A^k,\nn\\
        \bar G_i&=G_i-b_{ij}F^j+\qty(-\frac{1}{2g^2}+\frac{\alpha'}{4})\epsilon_{ijk}A^j\land A^k,\nn\\
        \bar G_{ij}&=Db_{ij}+\epsilon_{kij}B_k+\qty(g^{-2}-\frac{\alpha'}{2})\epsilon_{kij}A^k.
\end{align}
In order for $\bar G_i$ and $\bar G_{ij}$ to be proper gauge-covariant field strengths for $B_i$ and $b_{ij}$, we must choose the four-derivative truncation
\begin{equation}
        B_i=\qty(-g^{-2}+\frac{\alpha'}{2})A^i.
\end{equation}
which results in the modified field strengths truncating to
\begin{equation}
        \bar h=\dd b+\qty(g^{-2}-\frac{\alpha'}{2})\omega_{3Y},\qquad\bar G_i=\qty(-g^{-2}+\frac{\alpha'}{2})F^i,\qquad \bar G_{ij}=0,
\end{equation}
where we have defined the Yang-Mills Chern-Simons term as usual
\begin{equation}
        \omega_{3Y}=A^i\land F^i+\frac{1}{6}\epsilon^{ijk}A^i\land A^j\land A^k.
\end{equation}
One may be tempted then to view this as a shift of $g$
\begin{equation}
        \tilde g^{-2}= g^{-2}-\frac{\alpha'}{2}.\label{eq:gshift}
\end{equation}
Comparing with the field redefinitions of \cite{Liu:2023fqq} and shifting $g$ according to \eqref{eq:gshift}, we naturally infer the truncation
\begin{equation}
        \delta B_i=\frac{1}{2}A^i,\qquad\delta g_{ij}=\frac{1}{4g^4}F^i_{\alpha\beta} F^j_{\alpha\beta}-\frac{1}{2}\delta_{ij},\qquad \delta b_{ij}=0.
\label{eq:trunc}
\end{equation}
We will show that \eqref{eq:trunc} is indeed a consistent truncation. While we omit some details, the steps parallel those of \cite{Liu:2023fqq}, with additional terms due to the gauging that must be taken care of.

\subsubsection{Consistency of the truncation}

We start with the internal components of the Einstein equation.  Given (\ref{eq:trunc}), the shift to the two-derivative equations of motion for $g_{ab}$ are
\begin{align}
        \delta\mathcal E^{(0)}_{ab}=\,&-e^{2\varphi}\nabla^\gamma (e^{-2\varphi} \delta P_{\gamma ab})+2\delta P_{\gamma ad}Q_{\gamma db}-\frac{1}{6}\mathcal C_{f,ac}\delta\mathcal C_{f,bc}\nonumber\\
        &+\frac{1}{4}\qty(F_{\gamma\delta}^iF_{\gamma\delta}^j\delta(e_i^ae_j^b)-2\tilde G_{\gamma\delta i}\delta\tilde G_{\gamma\delta j}e^i_a e^j_b-\tilde G_{\gamma\delta i}\tilde G_{\gamma\delta j}\delta(e^i_a e^j_b))\nonumber\\
        &-\frac{m^2}{4}\epsilon_{ik\ell}\epsilon_{jmn}\delta(g^{km}g^{\ell n}e^i_a e^j_b),
\label{eq:dEab0}
\end{align}
where the $(ab)$ indices are implicitly symmetrized. Substituting in the lowest order equations of motion and making use of the $h$ Bianchi identity \eqref{eq:hbianchi}, one can derive the useful formula
\begin{align}
        e^{2\varphi}\nabla^\alpha\qty(e^{-2\varphi}\delta P_{\alpha ab})=&\Bigl[ -\frac{1}{4g^2}R_{\alpha\beta\gamma\delta}(\omega_+)F^i_{\alpha\beta}F^j_{\gamma\delta}+\frac{1}{4g^2}D_\gamma F_{\alpha\beta}^i D_\gamma F_{\alpha\beta}^j+\frac{1}{4g^4}F^i_{\alpha\beta}F^{k}_{\alpha\beta}F^j_{\gamma\delta}F^k_{\gamma\delta}\nn\\
        &+\frac{1}{2g^4} F^i_{\alpha\beta}F^j_{\beta\gamma}F^k_{\gamma\delta}F^k_{\delta\alpha}-\frac{1}{2g^4} F^i_{\alpha\beta}F^k_{\beta\gamma}F^j_{\gamma\delta}F^k_{\delta\alpha}-\frac{1}{2g^2} \epsilon^{jk\ell}F^i_{\alpha\beta}F^{k}_{\beta\gamma}F^\ell_{\gamma\alpha}\nn\\
        &-\frac{1}{2g}D_\gamma\mathcal E^{(0)}_{H,\delta j}F^i_{\gamma\delta}-\frac{1}{2g^2} \mathcal E^{(0)}_{g,\alpha\beta}F^j_{\gamma\alpha}F^i_{\beta\gamma}\Bigr]\delta^i_{(a}\delta^j_{b)}.
\end{align}
We may also evaluate
\begin{align}
        \mathcal E^{(1)}_{g,ab}=&\Bigg[-\frac{1}{8g^2}F^i_{\gamma\delta}F^j_{\alpha\beta}\qty(2R_{\gamma\delta}{}^{\alpha\beta}(\omega_+)-g^{-2}F^k_{\gamma\delta}F^k_{\alpha\beta})+\frac{1}{2g^4}F^i_{\alpha\beta}F^j_{\beta\gamma}F^k_{\gamma\delta}F^k_{\delta\alpha}-\frac{1}{2g^4}F^i_{\alpha\beta}F^k_{\beta\gamma}F^j_{\gamma\delta}F^k_{\delta\alpha}\nn\\
        &+\frac{1}{4g^2}D^{(+)}_\gamma F^i_{\alpha\beta}D^{(+)}_\gamma F^j_{\alpha\beta}+\frac{1}{2}\delta_{ij}-\frac{1}{2g^2}\epsilon^{ik\ell}F^k_{\alpha\beta}F^{\ell}_{\beta\gamma}F^j_{\gamma\alpha}+\frac{1}{4}F^2\delta_{ij}-\frac{1}{4}F^i_{\alpha\beta}F^j_{\alpha\beta}\nn\\
        &-\frac{1}{4g}\epsilon^{ik\ell}A^k_{\gamma}A^\ell_\delta F^i_{\gamma\delta}\Bigg]\delta^i_{(a}\delta^j_{b)},
\label{eq:scalarEOM1}
\end{align}
which may then be used to obtain that
\begin{equation}
        \delta\mathcal E^{(0)}_{g,ij}+\mathcal E^{(1)}_{g,ij}=\frac{1}{2g}\qty(D_\gamma\mathcal E^{(0)}_{H,\delta j}F^i_{\gamma\delta}+g^{-1} \mathcal E^{(0)}_{g,\alpha\beta}F^j_{\gamma\alpha}F^i_{\beta\gamma}),
\label{eq:scalarEOM}
\end{equation}
which vanishes upon imposing the leading-order equations of motion. In particular, the last term in \eqref{eq:scalarEOM1} with the bare $A$'s is precisely canceled by the corresponding shift $\delta\tilde G_i$ in (\ref{eq:dEab0}).

Similarly, $\delta\mathcal E^{(0)}_{H,ij}$ almost vanishes except for the bare $A^i$ that show up to account for the shifts to the modified field strengths $\bar G_i$ and $\bar G_{ij}$, but these are precisely canceled by the terms appearing in $\mathcal E^{(1)}_{H,ij}$. Keeping careful track of terms, it is straightforward to work out that
\begin{equation}
        \delta\mathcal E^{(0)}_{H,ij}+\mathcal E^{(1)}_{H,ij}=-g^{-1}D_\alpha\mathcal E^{(0)}_{H,\beta j}F^i_{\alpha\beta}-g^{-2}\mathcal E^{(0)}_{H,\alpha\beta}F^i_{\alpha\gamma}F^j_{\beta\gamma},
\end{equation}
which also vanishes by the leading-order equations of motion. This thus confirms that it is consistent to truncate out the scalars $g_{ij}$ and $b_{ij}$.
    
Finally, we turn our attention to the compatibility of the two Yang-Mills equations. Again, we can derive a useful formula
\begin{align}
    e^{2\varphi}D^\beta\qty(e^{-2\varphi}R_{\beta\alpha}{}^{\gamma\delta}(\omega_+)F^i_{\gamma\delta})=&\Bigg[-g^{-2}\nabla_\gamma\qty(F^{j}_{\delta\epsilon}F^{j}_{\epsilon\alpha})-\frac{1}{4}\nabla_\gamma\qty(h_{\delta\beta\epsilon}h_{\alpha\beta\epsilon})+\frac{1}{2}R_{\gamma[\delta|\beta\epsilon}h_{|\alpha]\beta\epsilon}\nn\\
    &-\frac{1}{2g^2}h_{\delta\alpha\epsilon}F^{j}_{\gamma\beta}F^{j}_{\epsilon\beta}-\frac{1}{8}h_{\delta\alpha\epsilon}h_{\gamma\beta\omega}h_{\epsilon\beta\omega}-\frac{1}{4g^2}h_{\alpha\beta\epsilon}F^{j}_{\beta\epsilon}F^{j}_{\gamma\delta}\nn\\
    &+\frac{1}{2g^2}F^{j}_{\alpha\beta}D^\beta F^{j}_{\gamma\delta}-\frac{1}{2g^2}h_{\delta\beta\epsilon}F^{j}_{\beta\epsilon}F^{j}_{\alpha\gamma}-g^{-2}F^{j}_{\beta\delta}D^\beta F^{j}_{\alpha\gamma}\nn\\
    &+\frac{1}{4}h_{\beta\epsilon\delta}\nabla^\beta h_{\alpha\gamma\epsilon}-\nabla_{[\gamma}\mathcal E^{(0)}_{g,\delta]\alpha}-\frac{1}{2}\nabla_\gamma\mathcal E^{(0)}_{H,\delta\alpha}-\partial_\gamma\varphi\mathcal E^{(0)}_{H,\delta\alpha}\nn\\
    &-\frac{1}{2}h_{\delta\alpha\epsilon}\mathcal E^{(0)}_{H,\gamma\epsilon}+\frac{1}{4}h_{\alpha\gamma\epsilon}\mathcal E^{(0)}_{H,\epsilon\delta}\Bigg]F^{i}_{\gamma\delta}+R_{\alpha\beta}^{\ \ \ \gamma\delta}(\omega_+)D^\beta F^{i}_{\gamma\delta}.
\end{align}
which may then be used to show that
\begin{align}
    &\qty(\delta \mathcal E^{(0)}_{H,\alpha a}+\mathcal E^{(1)}_{H,\alpha a})+2\qty(\delta \mathcal E^{(0)}_{g,\alpha a}+\mathcal E^{(1)}_{g,\alpha a})\nn\\
    &\kern4em=-\frac{1}{2}\delta_i^aF^{i}_{\gamma\delta}\Bigg[-\nabla_{[\gamma}\mathcal E^{(0)}_{g,\delta]\alpha}-\frac{1}{2}\nabla_\gamma\mathcal E^{(0)}_{H,\delta\alpha}-\partial_\gamma\varphi\mathcal E^{(0)}_{H,\delta\alpha}-\frac{1}{2}h_{\delta\alpha\epsilon}\mathcal E^{(0)}_{H,\gamma\epsilon}+\frac{1}{4}h_{\alpha\gamma\epsilon}\mathcal E^{(0)}_{H,\epsilon\delta}\Bigg]\nn\\
    &\kern5em+\delta_i^a\qty(\delta e^a_i \mathcal E^{(0)}_{H,\alpha a}+2\delta e^i_a\mathcal E^{(0)}_{g,\alpha a}),
\end{align}
which demonstrates that the two equations are indeed consistent after imposing the two-derivative equations of motion.

\subsubsection{The surviving equations of motion}

Here we summarize the equations of motion for the remaining degrees of freedom, namely the 7D metric $g_{\mu\nu}$, the two-form $b$-field, the graviphoton $A^i$, and the dilaton $\varphi$.  The equations of motion for the metric become
\begin{align}
    \mathcal E_{g,\alpha\beta}&=R(\omega)_{\alpha\beta}-\tilde g^{-2}F_{\alpha\gamma}^{i}F_{\beta\gamma}^{i}-\fft14\tilde h_{\alpha\gamma\delta}\tilde h_{\beta\gamma\delta}+2\nabla_\alpha\nabla_\beta\varphi\nn\\
    &\quad+\frac{\alpha'}{4}\biggl(R_{\alpha\gamma\delta\epsilon}(\omega_+)R_\beta{}^{\gamma\delta\epsilon}(\omega_+)-4g^{-2}R_{\alpha}{}^{\gamma\delta\epsilon}(\omega_+)F^{i}_{\beta\gamma}F^{i}_{\delta\epsilon}\nn\\
    &\kern4em+2g^{-4}F^{i}_{\alpha\gamma}F_{\beta}{}^{\gamma\,j}F^{i}_{\delta\epsilon}F^{\delta\epsilon\,j}+g^{-2}D^{(+)}_\alpha F^{i}_{\gamma\delta}D^{(+)}_
    \beta F^{\gamma\delta\,i}\biggr),
\end{align}
where $(\alpha\beta)$ symmetrization is assumed implicitly. Here, we have conveniently defined
\begin{equation}
        \tilde h\equiv\bar h-\frac{\alpha'}{4}\omega_{3L}(\omega_+)=\dd b+\tilde g^{-2}\omega_{3Y}-\frac{\alpha'}{4}\omega_{3L}(\omega_+),
\end{equation}
such that the Bianchi identity becomes
\begin{equation}
        \dd\tilde h=\tilde g^{-2}F^i\land F^i-\frac{\alpha'}{4}\Tr R(\omega_+)\land R(\omega_+).
\end{equation}
This newly defined $\tilde h$ then has the equation of motion
\begin{equation}
        \mathcal E_{H,\alpha\beta}=e^{2\varphi}\nabla^\gamma\qty(e^{-2\varphi}\tilde h_{\alpha\beta\gamma}).
\end{equation}

The graviphoton equation becomes
\begin{align}
    \mathcal E_{A,\alpha i}&=\tilde g^{-1}e^{2\varphi}\nabla^\gamma\qty(e^{-2\varphi}F_{\gamma\alpha}^{i})-\frac{1}{2\tilde g}\tilde h_{\alpha\beta\gamma}F_{\beta\gamma}^{i}\nn\\
    &\quad+\fft{\alpha'}4\biggl(-g^{-1}\tilde h_{\alpha\beta\gamma}R_{\beta\gamma\delta\epsilon}(\omega_+)F_{\delta\epsilon}^{i}+g^{-3}\tilde h_{\alpha\beta\gamma}F_{\beta\gamma}^{j}F_{\delta\epsilon}^{i}F_{\delta\epsilon}^{j}-2g^{-1}R_{\alpha\beta\gamma\delta}(\omega_+)D_\beta^{(+)}F_{\gamma\delta}^{i}\nn\\
    &\kern4em+2g^{-3}F_{\alpha\gamma}^{j}F_{\delta\epsilon}^{j}D_\gamma^{(+)}F_{\delta\epsilon}^{i}-2g^{-3}F_{\alpha\gamma}^{j}F_{\delta\epsilon}^{i}D_\gamma^{(+)}F_{\delta\epsilon}^{j}+4g^{-3}F_{\beta\gamma}^{i}F_{\gamma\delta}^{j}D_\alpha^{(+)}F_{\beta\delta}^{j}\nn\\
    &\kern4em +\frac{1}{4g}\epsilon^{ijk}F_{\gamma\delta}^jD^{(+)}_\alpha F^k_{\gamma\delta}\biggr),
\label{eq:redA}
\end{align}
while the dilaton equation becomes
\begin{align}
    \mathcal E_\phi&=R(\omega)-\frac{1}{2\tilde g^2}F_{\alpha\beta}^{i}F_{\alpha\beta}^{i}-\fft1{12}\tilde h_{\alpha\beta\gamma}^2+4\Box\varphi-4(\partial\varphi)^2+\tilde g^2\nn\\
    &\quad+\fft{\alpha'}8\biggl((R_{\alpha\beta\gamma\delta}(\omega_+))^2-4g^{-2}R_{\alpha\beta\gamma\delta}(\omega_+)F_{\alpha\beta}^{i}F_{\gamma\delta}^{i}+2g^{-2}\qty(D_\alpha^{(+)}F_{\beta\gamma}^{i})^2\nn\\
    &\kern4em+2g^{-4}F_{\alpha\beta}^{i}F_{\beta\gamma}^{i}F_{\gamma\delta}^{j}F_{\delta\alpha}^{j}-2g^{-4}F_{\alpha\beta}^{i}F_{\beta\gamma}^{j}F_{\gamma\delta}^{i}F_{\delta\alpha}^{j}+2g^{-4}F_{\alpha\beta}^{i}F_{\alpha\beta}^{j}F_{\gamma\delta}^{i}F_{\gamma\delta}^{j}\nn\\
    &\kern4em-\frac{1}{3g^2}\epsilon^{ijk}F^i_{\alpha\beta}F^j_{\beta\gamma}F^k_{\gamma\alpha}\biggr).
    \label{eq:reddil}
\end{align}

Having shown that the truncation is consistent, we may also compute the truncated Lagrangian to be
\begin{align}
    e^{-1}\mathcal L=&e^{-2\varphi}\Biggl[R(\omega)+4(\partial\varphi)^2-\fft1{12}\tilde h_{\alpha\beta\gamma}^2-\frac{1}{2\tilde g^2}\qty(F_{\alpha\beta}^{i})^2+\tilde g^2\nn\\
    &\kern2em+\fft{\alpha'}8\biggl((R_{\alpha\beta\gamma\delta}(\omega_+))^2-4\tilde g^{-2}R_{\alpha\beta\gamma\delta}(\omega_+)F_{\alpha\beta}^{i}F_{\gamma\delta}^{i}+2\tilde g^{-2}\qty(D_\alpha^{(+)}F_{\beta\gamma}^{i})^2\nn\\
    &\kern5em+2\tilde g^{-4}F_{\alpha\beta}^{i}F_{\beta\gamma}^{i}F_{\gamma\delta}^{j}F_{\delta\alpha}^{j}-2\tilde g^{-4}F_{\alpha\beta}^{i}F_{\beta\gamma}^{j}F_{\gamma\delta}^{i}F_{\delta\alpha}^{j}+2\tilde g^{-4}F_{\alpha\beta}^{i}F_{\alpha\beta}^{j}F_{\gamma\delta}^{i}F_{\gamma\delta}^{j}\nn\\
    &\kern5em-\frac{1}{3\tilde g^2}\epsilon^{ijk}F^i_{\alpha\beta}F^j_{\beta\gamma}F^k_{\gamma\alpha}\biggr)\Biggr].\label{eq:effectiveL}
\end{align}
Notably, all of the coupling constant $g$'s at the two-derivative level combine in just the right way with $\alpha'$ so as to be consistent with the shift \eqref{eq:gshift}.  Since we are working only to first order in $\alpha'$, we have also replaced $g$ by $\tilde g$ in the $\mathcal O(\alpha')$ contribution to the Lagrangian.

\section{The fermionic truncation}\label{sec:fermionic}

We now turn our attention to the fermion sector. The gravitino $\psi_M$ naturally splits into components along the spacetime directions $\psi_\mu$, which should be interpreted as the lower-dimensional gravitino, and components along the internal directions $\psi_i$, which should be interpreted as gaugini for the vectors $B_i$. Since we are truncating away the $B_i$, we expect to also truncate out the associated gaugini. 

Since we are reducing to seven dimensions, it is useful to decompose our gamma matrices as
\begin{align}
    \Gamma^\alpha&=\gamma^\alpha\otimes\mathbbm{1}\otimes\sigma^1,\nn\\
    \Gamma^a&=\mathbbm{1}\otimes\tau^a\otimes\sigma^2.
\end{align}
Here the $\gamma^\alpha$ form a seven-dimensional Clifford algebra $\mathrm{Cliff}(6,1)$, while the $\tau^a$ are the Pauli matrices of our three-dimensional Clifford algebra $\mathrm{Cliff}(3)$\footnote{We have denoted Pauli matrices by both $\tau^a$ and $\sigma^i$. While they are the same matrices, this is done to clarify which spinor subspace they are acting on.}. We take the convention that $\gamma^{0123456}=1$ and $\tau^{789}=i$. The chirality matrix then becomes
\begin{equation}
    \Gamma_{11}=\Gamma^{0123456789}=-\mathbbm{1}\otimes\mathbbm{1}\otimes\sigma^3.
\end{equation}
The choice of 10D chirality, which we take to be $\Gamma_{11}\epsilon=-\epsilon$, thus implies that $\sigma^3\epsilon=\epsilon$.  The ten-dimensional gravitino has the same chirality as $\epsilon$, while the ten-dimensional dilatino has the opposite.  Thus we can represent the heterotic Majorana-Weyl spinors as
\begin{equation}
    \epsilon\to\epsilon\otimes\begin{bmatrix}1\\0\end{bmatrix},\qquad
    \psi_M\to\psi_M\otimes\begin{bmatrix}1\\0\end{bmatrix},\qquad
    \lambda\to\lambda\otimes\begin{bmatrix}0\\1\end{bmatrix}.
\end{equation}
The spinors on the right-hand side of these expressions are 16 component spinors that further decompose into a pair of seven-dimensional spinors that are acted on by $\tau^a$.  This pair of spinors satisfies a Majorana condition that we do not concern ourselves with here.

After truncation, the leading order supersymmetry variations become
\begin{align}
        \delta_\epsilon\psi_\mu^{(0)}=&\qty[D_\mu(\omega_-)+\frac{i}{2g}F_{\mu\nu}^i\gamma^\nu\tau^{\underline i}]\epsilon,\nn\\
        \delta_\epsilon\psi_i^{(0)}=&\,0,\nn\\
        \delta_\epsilon\tilde\lambda^{(0)}=&\qty[\gamma^\mu\partial_\mu\varphi-\frac{1}{12}h_{\mu\nu\rho}\gamma^{\mu\nu\rho}+\frac{i}{4g}F^i_{\mu\nu}\gamma^{\mu\nu}\tau^{\underline i}-\frac{g}{2}]\epsilon.
\end{align}
It is also noteworthy that the composite $SU(2)$ connection, upon truncation, becomes the \textit{gauge} $SU(2)$ connection
\begin{equation}
        Q_{\alpha ab}=\epsilon^{ijk}\delta^i_a\delta^j_b A^k_\alpha,
\end{equation}
which is what promotes the covariant derivative to a gauge-covariant one in $\delta_\epsilon\psi_\mu$, which acts by
\begin{equation}
        D_\mu\epsilon=\nabla_\mu\epsilon+\frac{i}{2}A^i_\mu\tau^{\underline{i}}\epsilon.
\end{equation}
In analogy to the torus case \cite{Liu:2023fqq}, we make the definition
\begin{equation}
        \mathcal D_\mu\equiv D_\mu(\omega_-)+\frac{i}{2g}F_{\mu\nu}^i\gamma^\nu\tau^{\underline i},
\end{equation}
which will be useful for the fermionic field redefinitions.

The bosonic field redefinitions \eqref{eq:trunc} combined with the two-derivative truncation \eqref{eq:truncation} lead to four-derivative contributions to the supersymmetry variations
\begin{align}
        \delta\qty(\delta_\epsilon\psi_\mu^{(0)})+\delta_\epsilon\psi^{(1)}_\mu=&\frac{1}{32}\Bigg[\omega_{3L,\mu\nu\rho}(\omega_+)\gamma^{\nu\rho}+4ig^{-1}\qty(R_{\mu\nu}{}^{\alpha\beta}(\omega_+)-\frac{1}{2g^2}F^{j}_{\mu\nu}F^{j}_{\alpha\beta})F_{\alpha\beta}^{i}\gamma^\nu\tau^{\underline i}\nn\\
        &\qquad+2g^{-2}F^{i}_{\alpha\beta}D^{(+)}_\mu F^{j}_{\alpha\beta}\tau^{\underline{ij}}-4igF^i_{\mu\nu}\gamma^\nu\tau^{\underline i}\Bigg]\epsilon,\nn\\
        \delta\qty(\delta_\epsilon\psi_i^{(0)})+\delta_\epsilon\psi^{(1)}_i=&\fft1{16g}F_{\alpha\beta}^{i}\Big(R_{\gamma\delta}{}^{\alpha\beta}(\omega_+)\gamma^{\gamma\delta}-g^{-2}F_{\alpha\beta}^{j}F_{\gamma\delta}^{j}\gamma^{\gamma\delta}-2ig^{-1}D_\gamma^{(+)} F_{\alpha\beta}^{j}\gamma^\gamma\tau^{\underline j}\nn\\
        &\qquad\qquad-2g^{-2}F_{\beta\gamma}^{j}F_{\gamma\alpha}^{k}\tau^{\underline{jk}}-2iF^j_{\alpha\beta}\Gamma^{\underline j}\Big)\epsilon,\nn\\
        \delta\qty(\delta_\epsilon\tilde\lambda^{(0)})+\delta_\epsilon\tilde\lambda^{(1)}=&\frac{1}{48}\biggl[\omega_{3L,\mu\nu\rho}(\omega_+)\gamma^{\mu\nu\rho}+3ig^{-1}\left(R_{\mu\nu}{}^{\alpha\beta}(\omega_+)-\frac{1}{2g^2}F_{\mu\nu}^{j}F_{\alpha\beta}^{j}\right)F_{\alpha\beta}^{i}\gamma^{\mu\nu}\tau^{\underline{i}}\nn\\
        &\kern2em-2g^{-3}\epsilon^{ijk}F^{i}_{\alpha\beta}F^{j}_{\beta\gamma}F^{k}_{\gamma\alpha}-6g^3-3igF^i_{\mu\nu}\gamma^{\mu\nu}\tau^{\underline i}-\frac{6}{g}F^i_{\alpha\beta}F^i_{\alpha\beta}\biggr]\epsilon.
\label{eq:susyvar}
\end{align}
While the supersymmetry variation of the gaugino, $\delta_\epsilon\psi_i$, is undesirable, it has but a single extra term compared to the ungauged case \cite{Liu:2023fqq}. In particular, using the fact that
\begin{equation}
        [D_\mu,D_\nu]\epsilon=\frac{1}{4}\qty(R_{\mu\nu}{}^{\alpha\beta}\gamma^{\alpha\beta}+\epsilon^{ijk}F_{\mu\nu}^i\Gamma^{\underline{jk}})\epsilon,
\end{equation}
we see that an analogous field redefinition holds as that in \cite{Liu:2023fqq}
\begin{equation}
        \tilde\psi_i=\psi_i-\frac{\alpha'}{2g}F^i_{\mu\nu}\mathcal D_\mu\psi_\nu,
\end{equation}
such that
\begin{equation}
        \delta_\epsilon\tilde\psi_i=0.
\end{equation}
Thus truncating $\tilde\psi_i$ is indeed consistent with supersymmetry.

Interestingly, the higher-derivative corrections in the variations, (\ref{eq:susyvar}), appear in exactly the appropriate way to be consistent with the shifted gauge coupling \eqref{eq:gshift}. This is more readily seen in the combined expressions
\begin{align}
    \delta_\epsilon\psi_\mu=&\biggl[D_\mu(\tilde\omega_-)+\frac{i}{2\tilde g}F_{\mu\nu}^i\gamma^\nu\tau^{\underline i}+\fft{i\alpha'}{8\tilde g}\qty(R_{\mu\nu}{}^{\alpha\beta}(\omega_+)-\frac{1}{2\tilde g^2}F^{j}_{\mu\nu}F^{j}_{\alpha\beta})F_{\alpha\beta}^{i}\gamma^\nu\tau^{\underline i}\nn\\
    &\qquad+\fft{\alpha'}{16\tilde g^{2}}F^{i}_{\alpha\beta}D^{(+)}_\mu F^{j}_{\alpha\beta}\tau^{\underline{ij}}\biggr]\epsilon,\nn\\
    \delta_\epsilon\tilde\lambda=&\biggl[\gamma^\mu\partial_\mu\varphi-\frac{1}{12}h_{\mu\nu\rho}\gamma^{\mu\nu\rho}+\frac{i}{4\tilde g}F^i_{\mu\nu}\gamma^{\mu\nu}\tau^{\underline i}-\frac{\tilde g}{2}-\frac{\alpha'}{8\tilde g}F^i_{\alpha\beta}F^i_{\alpha\beta}\nn\\
    &\kern2em+\fft{i\alpha'}{16\tilde g}\left(R_{\mu\nu}{}^{\alpha\beta}(\omega_+)-\frac{1}{2\tilde g^2}F_{\mu\nu}^{j}F_{\alpha\beta}^{j}\right)F_{\alpha\beta}^{i}\gamma^{\mu\nu}\tau^{\underline{i}}-\fft{\alpha'}{24\tilde g^3}\epsilon^{ijk}F^{i}_{\alpha\beta}F^{j}_{\beta\gamma}F^{k}_{\gamma\alpha}\biggr]\epsilon,
\end{align}
where again we make no distinction between $g$ and $\tilde g$ in the $\mathcal{O}(\alpha')$ terms.

\section{Discussion}\label{sec:discussion}
In this paper, we have shown that the SU(2) group manifold reduction of four-derivative heterotic supergravity on $S^3$ may be consistently truncated to pure half-maximal gauged 7D supergravity. This may be seen as supporting evidence that the Gauntlett-Varela conjecture \cite{Gauntlett:2007ma} extends to higher-derivative truncations. 

We may, of course, compare our results to those of the ungauged case \cite{Liu:2023fqq}. This requires that we rescale $A^i\to g A^i$ so as to recover the proper normalization and also rescale $\sigma^i\to g\sigma^i$ so that the Maurer-Cartan equation becomes
\begin{equation}
    \dd\sigma^i=-\frac{g}{2}\epsilon^{ijk}\sigma^j\land\sigma^k.
\end{equation}
Upon sending $g\to 0$, the field strength becomes abelian and the $\sigma^i$ become torus coordinates
\begin{equation}
   F^i=\dd A^i,\qquad \dd \sigma^i=0.
\end{equation}
This also has the effect of reducing the gauge-covariant derivatives $D$ to regular covariant derivatives $\nabla$. Writing $\sigma^i=\dd y^i$, we then recover the torus metric
\begin{equation}
    \dd s^2=g_{\mu\nu}\dd x^\mu\dd x^\nu+(\dd y^i+ A^i)^2,
\end{equation}
as well as the ungauged Lagrangian
\begin{align}
        e^{-1}\mathcal L=&e^{-2\varphi}\Biggl[R+4(\partial\varphi)^2-\fft1{12}\tilde h_{\alpha\beta\gamma}^2-\fft12\qty(F_{\alpha\beta}^{i})^2\nn\\
    &\kern2em+\fft{\alpha'}8\biggl((R_{\alpha\beta\gamma\delta}(\omega_+))^2-4R_{\alpha\beta\gamma\delta}(\omega_+)F_{\alpha\beta}^{i}F_{\gamma\delta}^{i}+2\qty(\nabla_\alpha^{(+)}F_{\beta\gamma}^{i})^2\nn\\
    &\kern5em+2F_{\alpha\beta}^{i}F_{\beta\gamma}^{i}F_{\gamma\delta}^{j}F_{\delta\alpha}^{j}-2F_{\alpha\beta}^{i}F_{\beta\gamma}^{j}F_{\gamma\delta}^{i}F_{\delta\alpha}^{j}+2F_{\alpha\beta}^{i}F_{\alpha\beta}^{j}F_{\gamma\delta}^{i}F_{\gamma\delta}^{j}\biggr)\Biggr].
    \end{align}
Moreover, since the shift $\sigma\to g\sigma$, $A^i\to g A^i$ effectively rescales $g_{ij}$ by $g^2$, the internal metric becomes that of the torus case
\begin{equation}
   g_{ij}=\delta_{ij}+\frac{\alpha'}{4}F^i_{\alpha\beta}F^j_{\alpha\beta}.
\end{equation}
We also recover the ungauged supersymmetry variations (in the original ten-dimensional Dirac matrix notation)
\begin{align}
    \delta_\epsilon\psi_\mu=&\Bigg[D_\mu(\tilde\omega_-)+\frac{1}{2}F_{\mu\nu}^i\Gamma^\nu\Gamma^{\underline i}+\frac{\alpha'}{8}\qty(R_{\mu\nu}{}^{\alpha\beta}(\omega_+)-\frac{1}{2}F^{j}_{\mu\nu}F^{j}_{\alpha\beta})F_{\alpha\beta}^{i}\Gamma^\nu\Gamma^{\underline i}\nn\\
    &\ +\frac{\alpha'}{16}F^{i}_{\alpha\beta}\nabla^{(+)}_\mu F^{j}_{\alpha\beta}\Gamma^{\underline{ij}}\Bigg]\epsilon,\nn\\
    \delta_\epsilon\tilde\lambda=&\Bigg[\Gamma^\mu\partial_\mu\varphi-\frac{1}{12}\tilde h_{\mu\nu\rho}\Gamma^{\mu\nu\rho}+\frac{1}{4}F^i_{\mu\nu}\Gamma^{\mu\nu}\Gamma^{\underline i}+\fft{\alpha'}{16}\left(R_{\mu\nu}{}^{\alpha\beta}(\omega_+)-\frac{1}{2}F_{\mu\nu}^{j}F_{\alpha\beta}^{j}\right)F_{\alpha\beta}^{i}\Gamma^{\mu\nu}\Gamma^{\underline{i}}\nn\\
    &\ +\fft{\alpha'}{24}F^{i}_{\alpha\beta}F^{j}_{\beta\gamma}F^{k}_{\gamma\alpha}\Gamma^{\underline{ijk}}\Bigg]\epsilon,
\end{align}
where
\begin{equation}
    \tilde\omega_-\equiv\omega-\frac{1}{2}\tilde h=\omega-\frac{1}{2}h+\frac{\alpha'}{8}\omega_{3L}.
\end{equation}
Hence, our results are consistent with \cite{Liu:2023fqq}, as they should be.

It is also interesting to note that the terms proportional to $g$ at the two-derivative level always have corresponding four-derivative terms in exactly the way so as to respect the shift \eqref{eq:gshift}.  While we have restricted our attention to $S^3$, there is no reason to suspect that this is specific to that setup. In particular, assuming there are no other obstructions to truncation, we should generically expect \cite{Lu:2006ah}
\begin{align}
    H_{ijk}=mf_{ijk},
\end{align}
for a group manifold reduction on a unimodular Lie group  $G$, whose Lie algebra has structure constants $f^{ijk}$. In general, we expect the two-derivative truncation to be $m=-g^{-2}$ and $g_{ij}=g^{-2}\kappa_{ij}$, where $\kappa$ is the Cartan-Killing metric. Under the assumption that $G$ is compact and semi-simple, $\kappa$ is necessarily symmetric, non-degenerate, and positive-definite, so we write $\kappa_{ij}=\delta_{ab}k_i^a k_j^b$. The torsionful spin connection should, in general, be
\begin{equation}
    \Omega_+^{ab}=-f_{ijk}k^i_a k^j_b\sigma^k,
\end{equation}
which leads to the terms
\begin{equation}
    \omega_{3L,abc}\supset-2k_a^{i}k_b^j k_c^{k}f_{ijk},
\end{equation}
in the Lorentz-Chern-Simons form. This then \emph{suggests} that we ought always to get the effective coupling shift \eqref{eq:gshift} for a group manifold reduction of heterotic supergravity.

While we have truncated away the heterotic gauge fields right from the start, one might wonder if they may be included in the reduction.  In the current context, since we were interested in truncating out all the vector multiplets, it was natural to truncate them in 10D.  As this is a truncation to gauge singlets, the initial removal of the heterotic gauge fields is guaranteed to be consistent, even at the higher derivative order.  Nevertheless, it would be interesting to see how these play into the story of higher-derivative consistent truncations, and if they might obstruct truncation more generally. We leave this to future work.

It would be interesting to see how our results extend to more general group manifold or coset reductions. In particular, it is known that any unimodular Lie group $G$ may be used for a group manifold reduction of heterotic supergravity \cite{Lu:2006ah} and one is free (at least at the two-derivative level) to truncate out the vector multiplets that arise. One could then wonder if any new constraints on $G$ arise at the four-derivative level. It would also be interesting to see if consistency extends to more general coset reductions such as the $SO(4)/SO(3)$ coset reduction of heterotic supergravity \cite{Cvetic:2000dm}.

Finally, one may ask whether such higher derivative truncations may be done more systematically in the framework of Double Field Theory (DFT), along the lines of \cite{Eloy:2020dko}. In particular, gauged DFT was used in \cite{Baron:2017dvb} to construct a large class of consistent truncations, including the $S^3$ group manifold reduction. Indeed, after truncating the result in \cite{Baron:2017dvb} and performing suitable field redefinitions, one finds that the effective action matches \eqref{eq:effectiveL} as expected. In the process, one must truncate the $O(3,3)$ covariant packaging down to a subsector. For example, the truncation sets half of the components of the $O(3,3)$ field strength $\mathcal F^I=(F^{(+)\,i},F^{(-)\,i})$ to zero. More generally, while DFT can be extremely useful in constructing consistent truncations where all the fields in an $O(d,d)$ multiplet are kept, we still have to break apart the $O(d,d)$ covariant language to truncate away some of the multiplets and check the consistency with the equations of motion.

\section*{Acknowledgements}
This work was supported in part by the U.S. Department of Energy under grant DE-SC0007859. RJS was funded in part by a Leinweber Graduate Summer Fellowship and in part by a Rackham One-term Dissertation Fellowship.

\bibliographystyle{JHEP}
\bibliography{cite}
\end{document}